\documentstyle[preprint,aps]{revtex}
\begin{document}
\tightenlines
\preprint{Preprint ``Notas de F\'\i sica'' CBPF-NF-013/02.}
\title{Non BPS topological defect associated with two coupled real field}
\author { E. L. da Gra\c{c}a$^{\dagger}$ and R. de Lima Rodrigues\thanks{
Permanent address: Departamento de Ci\^encias Exatas e da Natureza,
Universidade Federal de Campina Grande,
Cajazeiras -- PB, 58.900-000, Brazil. E-mail: rafaelr@cbpf.br.}
\\
Centro Brasileiro de Pesquisas F\'\i sicas\\
Rua Dr. Xavier Sigaud, 150\\
CEP 22290-180, Rio de Janeiro-RJ, Brazil\\
${}^{\dagger}$ Departamento de F\'\i sica,
Universidade Federal Rural do Rio de Janeiro\\
Antiga Rodovia Rio-S\~ao Paulo Km 47, BR 465,
CEP 23.890-000, Serop\'edica-RJ
}

\maketitle

\begin{abstract}
 We investigate a stability equation involving two-component
eigenfunctions which is associated with  a 
potential model in terms of two coupled real scalar fields,
which presents non BPS topological defect.
\end{abstract}

\vspace{0.5cm}

\vspace{2cm}

This work was presented in the XXII Brazilian National
Meeting on Particles and Fields (October/2001), 
to appear into the site www.sbf.if.usp.br.

\pacs
\newpage

\section{Introduction}

We  consider the non Bogomol'nyi
\cite{Bogo} and Prasad-Sommerfield \cite{PS} (non BPS) 
classical soliton (defect) solutions 
of two coupled real scalar
fields. These are static, nonsingular classically stable
solutions of the field equations with finite localized energy
\cite{Jac,Raja,Cole,Bala}. Here we consider  a particular
case of  two-coupled-field systems, independent of the time
\cite{Raja,Boya}, so that they are non BPS soliton solutions. 
Such  static classical configurations in 1+1 dimensions we have 
kinks, in 2+1 dimensions the vortices and in 3+1 dimensions
the monopoles, the strings, and domain walls, etc. 

Recently the case of BPS
topological defect associated with two coupled real scalar fields  
in a system containing up to sixth-order powers in the fields
has been investigated
via supersymmetry in quantum mechanics \cite{RPV98}.
In this work, in the asymptotic region, we find
the zero mode eigenfunction associated with the stability
equation of a relativistic system 
containing up to fourth-order powers in the fields without $N=1$ supersymmetry.

Recently, one-loop quantum corrections
to soliton energies and central charges in the supersymmetric $\phi^4$ and
sine-Gordon models in (1+1)-dimensions have been investigated
\cite{Graha}.
The reconstruction of 2-dimensional scalar field potential models has
been considered and quantum corrections to the solitonic
sectors of both potentials are pointed out \cite{GN}.

\section{Topological defects for two coupled scalar fields}

The topological defects considered in this work are called kinks. 
Our potential model of two coupled
real scalar fields in 1+1 dimensions  presents $Z(2)$ symmetry
and the classical soliton solutions.

The Lagrangian density for such a nonlinear system in the natural system of
units $(c=\hbar=1)$, is given by

\begin{equation}
\label{E18}
{\cal L}\left(\phi, \chi, \partial_{\mu}
\phi, \partial_{\mu}\chi\right)
= \frac{1}{2}\left(\partial_{\mu}\phi\right)^2+
\frac{1}{2}\left(\partial_{\mu}\chi\right)^2 -V(\phi, \chi),
\end{equation}
where $\partial_{\mu}=\frac{\partial}{\partial x^{\mu}},
\quad x^{\mu}=(t,x) $ with $ \mu =0, 1,
\quad x_\nu=\eta_{\nu\mu}x^{\mu};
\quad
\phi =\phi (x,t), \quad\chi =\chi (x,t)$
are real scalar fields, and
$\eta^{\mu\nu}$ is the metric tensor given by

\begin{equation}
\label{E018}
\eta^{\mu\nu}=\left(
\begin{array}{cc}
1 & 0 \\
0 & -1
\end{array}\right).
\end{equation}
Here the potential $V=V(\phi,\chi)$ is given by

\begin{equation}
V(\phi, \chi)=\frac{\lambda}{4}\left(\phi^2-\frac{m^2}
{\lambda}\right)^2+\frac{\lambda}{4}
\left(\chi^2-\frac{\mu}{\lambda}\right)^2
+\gamma\chi^2\phi^2, 
\label{EP}
\end{equation}
 where $\lambda>0, m>0, \mu>0.$ Observe that this potential has a discrete
symmetry as $\phi\rightarrow-\phi$ and $\chi\rightarrow-\chi$ so that we
have a necessary condition that it must have at least
two zeroes in order that solitons can exist. However, this condition is not 
sufficient.

 The general classical
configurations obey the following equations of motion:

\begin{equation}
\label{EM}
 \frac{\partial^2}{\partial t^2}\phi-
\frac{\partial^2}{\partial x^2}\phi
+\frac{\partial }{\partial \phi}V=0, \qquad
 \frac{\partial^2}{\partial t^2}\chi-
\frac{\partial^2}{\partial x^2}\chi
+\frac{\partial }{\partial \chi}V=0,
\end{equation}
which, for static soliton solutions, become a
system of nonlinear differential equations given by

\begin{eqnarray}
\label{EME}
 \phi^{\prime\prime}&&=\frac{\partial }{\partial \phi}V=
\lambda\phi\left(\phi^2-m^2+2\gamma\chi^2\right)\nonumber\\
 \chi^{\prime\prime}&&=\frac{\partial }{\partial \chi}V=
\lambda\chi\left(\chi^2-\mu^2+2\gamma\phi^2\right),
\end{eqnarray}
where primes represent differentiations with respect to the space variable.
There are two vacua given by

\begin{equation}
M_i=\left(\pm\frac{m}{\sqrt{\lambda}}, 0\right), \quad
N_i=\left(0, \pm\frac{m}{\sqrt{\lambda}}\right).
\label{EVac}
\end{equation}
However, $M_i$ or $N_i$ is a false vacuum. Choosing the vacuum state
so that $\mu^2 <\nu^2,$ turns  $M_i$  a true vacuum given by

\begin{eqnarray}
M_1=&&\left(\frac{m}{\sqrt{\lambda}}, 0\right), 
\quad x\rightarrow\infty\nonumber\\
M_2=&&=\left(-\frac{m}{\sqrt{\lambda}}, 0\right), 
\quad x\rightarrow-\infty. 
\label{E2Vac}
\end{eqnarray}
We see that these solitons are non BPS because the Bogomol'nyi
form of the energy is not consisting of a sum of squares and the surface
term, i.e.

\begin{equation}
E_B\neq\int dx\frac{\partial}{\partial x}\Gamma [\phi(x), \chi(x)].
\end{equation}
The conserved topological current ($\partial_{\mu}j^{\mu}=0$) can not
be written in terms of the continuously twice differentiable function
$\Gamma(\phi, \chi)$, viz.,

\begin{equation}
\label{E31} 
j^{\mu}\neq\epsilon^{\mu\nu}\partial_{\nu}\Gamma (\phi, \chi),
\quad \epsilon^{00}=\epsilon^{11}=0,\quad \epsilon^{10}=-\epsilon^{01}=-1.
\end{equation}
Therefore, there is no superpotential $\Gamma(\phi, \chi)$ 
and, so, the topological charge of such a system is not equivalent to the
minimum value of the energy. Next we consider the particular case so  that
 $\gamma=\lambda^2$. The case with $\gamma\neq\lambda^2$ will be consider 
 elsewhere \cite{ger02a}.

Now let us analyze the connection between an approach for $\phi^4$ model
according Rajaraman's method so that we choose an ellipse orbit in the
plane $(\phi, \chi),$ viz.,

\begin{equation}
\chi^2(x)=\beta^2\left(1-\frac{\phi^2(x)}{\alpha^2}\right)
 \label{EE}
\end{equation}
where $\alpha$ and $\beta$ are real constants. 
Observe that in this trajectory as $x\rightarrow\infty, \phi\rightarrow
\phi_{vacuum}=\alpha$ because $\chi=0.$
From (\ref{EE}) and (\ref{EME}) and making  $\phi=\alpha\psi,$
we get the following Riccati equation

\begin{equation}
\label{ERC} 
\psi^{\prime\prime}(x)=2(\mu^2-m^2)[\psi^3(x)-\psi(x)], 
\end{equation}
which provides a particular solution given by

\begin{equation}
\label{ESRC} 
\psi(x)=\tanh[(\mu^2-m^2) x].
\end{equation}
This Riccati equation is a particular case of field equations in
scalar potential models in 1+1 dimensions \cite{Burt}.
In similar way we also have a Riccati equation associated to field $\chi$
which has a solution that satisfies the above orbit given in Eq.
(\ref{EE}). Therefore, we find the following static classical configurations

\begin{eqnarray}
\phi(x)=\alpha\tanh[(\mu^2-m^2) x], \quad
\alpha=\frac{m}{\sqrt{\lambda}}\nonumber\\
\chi(x)=\beta sech[(\mu^2-m^2) x], \quad 
\beta=\sqrt{2(3m^2-\mu^2)}.
\label{EK}
\end{eqnarray}

The classical stability of the
kinks in this non-linear system \cite{Boya,RPV98} is 
analyzed by considering small perturbations around $\phi (x)$ and 
$\chi(x)$:

\begin{equation}
\label{E32}
  \phi(x,t)=\phi(x)+\eta(x,t)
\end{equation}
and

\begin{equation}
\label{E33}
  \chi(x,t)=\chi(x)+\rho(x,t).
\end{equation}
Next let us  expand the fluctuations $\eta(x,t)$
and $\xi(x,t)$ in terms of the normal modes:

\begin{equation}
\label{E34}
\eta (x,t) = \sum_n \epsilon_n \eta_n (x) e^{i\omega_n t}
\end{equation}
and

\begin{equation}
\label{E35} \rho (x,t) = \sum_n c_n \rho_n (x) e^{i\tilde{\omega}_n t}.
\end{equation}
Thus, if $\tilde{\omega}_n=\omega_n,$ the equations of motion 
for the two fields become from (\ref{EM}) a Schr\"odinger-like
equation for two-component wave functions $\Psi_{n}$. If
 $\tilde{\omega}_n\neq\omega_n$ we obtain

\begin{equation}
\label{EED}
{\cal H}\Psi_{n}
= \tilde{\Psi}_{n}, \quad n=0, 1, 2, ...,
\end{equation}
where the fluctuation operator can be written as

\begin{equation}
\label{E37}
{\cal H}=\left(
\begin{array}{cc}
-\frac{d^2}{dx^2} +V_{11} &
V_{12} \\
V_{21} & -\frac{d^2}{dx^2} + V_{22}
\end{array}\right)_{\vert\phi =\phi(x), \chi =\chi(x) }
\end{equation}
with

\begin{eqnarray}
\label{define}
V_{11}=&&m^2\{3tanh^2[(\mu^2-m^2) x]-1\}\nonumber \\
V_{21}=&&V_{12}=4\gamma\alpha\beta\tanh[(\mu^2-m^2) x]
sech[(\mu^2-m^2) x]\nonumber \\
V_{22}=&&(-\mu^2+2m^2\lambda)\tanh^2[(\mu^2-m^2) x]
\end{eqnarray}
and

\begin{equation}
\label{E32a}
\tilde{\Psi}_{n}=\left(
\begin{array}{cc}
\omega_n^2\eta_n(x) \\
\tilde{\omega}_n^2 \rho_n(x)
\end{array}\right).
\end{equation}

When the two-component normal modes in (\ref{EED}) satisfy $\omega_n{^2}
\geq 0$ and $\tilde{\omega}_n^2\geq 0$  we have ensured  the stability 
of the defects.

We can calculate the two-component eigenfunction associated to  the zero mode
via classical analysis of second order differential equation. 
Here we present only the results for a particular case.
Making an extension for the case of
only one single real scalar field we can realize, {\it a priori}, the
asymptotic $(x\rightarrow\infty)$ behavior  
so that we obtain the following zero mode eigenfunction
($\tilde\omega=\omega=0)$

\begin{equation}
\label{EG1}
 \Psi_0(x)=C_1\left(
\begin{array}{c}
e^{-bx}\\
cos(\kappa x+\theta)
\end{array}\right),
\end{equation}
where $\theta$ is a real constant and $C_1$ is the normalization
constant.
As $m^2<\mu^2$ implies $\kappa^2=
2\frac{m^2}{\lambda}-\mu^2<0$, for $\lambda\geq 2.$
Besides, our results for kinks are ensured when $\mu$ and $\nu$ are
in the following interval: $\frac 13\leq\frac{m^2}{\mu^2}<1.$

\section{Conclusions}

The connection between the non-relativistic quantum mechanics 
with two-component wave functions
and the stability equations associated with defect (soliton) 
solutions for a model of
two coupled real scalar fields in 1+1 dimensions has been presented. In
this work, we have considered  an application for a potential associated
with the $\phi^4$ model in 1+1 dimensions which,
 was solved by the trial orbit method treated in the Ref. \cite{Raja}.

We see that, if $\Psi^{(0)}_-=\Psi_0(x)$ is a normalizable
two-component eigenstate of the bosonic sector, one cannot write 
$\Psi^{(0)}_+$ of the fermionic sector in terms of
$\Psi^{(0)}_-$ in a similar manner to ordinary supersymmetric quantum
mechanics. This fact can be checked in the example treated here.
A detailed analysis of our work will be published elsewhere.

Static classical field configurations
have been recently exploited
into the context of defect that live inside 
topological defects \cite{DF00}, and triple junctions via $N=1$ supersymmetry 
theories \cite{carro}
and without supersymmetry providing  networks of
domain wall \cite{DF}, and using the 
vinicity of BPS bound states\cite{Carlos01}. Recently the BPS saturated objects
with axial geometry (wall junctions, vortices), in generalized Wess-Zumino
 models have been investigated \cite{ShifV}.

We are investigating the insue of the stability, via
arguments based on supersymmetric quantum mechanics, for the BPS 
and non-BPS states from two domain walls in a potential model,
in boths situations,  without supersymmetry and in the case of a minimal
(N=1-SUSY) \cite{ger02a}.

\acknowledgments

We wish to thank the staff of the CBPF and CFP-UFCG of Cajazeiras-PB,
Brazil.  We would also like to thank J. A. Hela\"yel-Neto
and Rubens L. P. G. Amaral for many stimulating discussions. 
RLR would also like to thank the CNPq-Brazil by a post-doctoral fellowship.

\end{document}